\documentclass{mem}
\usepackage{natbib}\usepackage{txfonts}\usepackage{balance}
\usepackage{graphicx}
\usepackage[a4paper]{hyperref}
\idline{79}{1}
\begin{document}
\def\teff{$T\rm_{eff }$}
\def\kms{$\mathrm {km s}^{-1}$}

\title{
Seyfert Galaxies in the Local Universe: Analysis of Spitzer Spectra of a Complete Sample
}

   \subtitle{}

\author{
S. \,Tommasin\inst{1}, 
L.\, Spinoglio\inst{1},
K.\, Hainline\inst{2}
M. A.\, Malkan\inst{2}
\and H.\, Smith\inst{3}        }
          
  \offprints{S. Tommasin}
 
\institute{
Istituto di Fisica dello Spazio Interplanetario, INAF, Via Fosso del Cavaliere 100, I-00133 Roma, Italy
\email{Silvia.Tommasin@ifsi-roma.inaf.it, Luigi.Spinoglio@ifsi-roma.inaf.it}
\and
Astronomy Division, University of California, Los Angeles, CA 90095-1547, USA
\email{kevinhainline@astro.ucla.edu, Malkan@astro.ucla.edu}
\and 
Harvard-Smithsonian Center for Astrophysics, 60 Garden Street, Cambridge, MA 02138
\email {hsmith@cfa.harvard.edu} 
}

\authorrunning{Tommasin et al.}

\titlerunning{Spitzer spectra of Seyfert galaxies}

\abstract{
The Spitzer high resolution spectra of 72 Seyfert galaxies from the 12$\mu$m Galaxy Sample are presented and discussed. The presence of starburst components in these galaxies can be quantified by powerful mid-IR diagnostics tools (i.e. 11.25$\mu$m PAH feature equivalent width and the H$_2$ emission line intensity), as well as the AGN dominance can be measured by specific fine structure line ratios (e.g. [NeV]/[NeII], [NeV]/[SiII], etc.). The two types of Seyfert galaxies do not show any statistical difference in our diagnostic tools. However, the Seyfert 2's showing hidden Broad Line Regions in spectro-polarimetric observations have on average an higher AGN dominance, a weaker star formation component and a warmer [60 - 25] spectral index than those without broad emission lines. 

}
\maketitle{}

\section{Introduction}

From the observational evidence that the main energy-generating mechanisms in galaxies are black hole accretion and star formation and that Starbursts and AGNÕs may be linked in an evolutionary sequence, we present mid-infrared spectroscopy of a complete flux-limited sample of Seyfert galaxies with the aim to derive the bi-variate AGN and Star Formation luminosity functions in the Local Universe. 
Mid-infrared spectra can  provide a census of the dominance of the two processes at zero redshift and it will be the basis for any future comparison with the history of the energy production mechanisms along galaxy evolution as derived from future cosmological surveys.
Because most of the local active galaxies contain both an AGN and a starburst component, often obscured by dust, infrared spectroscopy has to be used to separate the two emission processes. Many spectroscopic indicators such as 
the ratios of high to low ionization lines, e.g.  [NeV]/[NeII], [NeV]/[SiII], [OIV]/[NeII], [NeIII]/[NeII] are directly linked to the AGN dominance, while others are indicators of the Star Formation dominance, e.g. the PAH 11.25$\mu$m, the molecular hydrogen rotational lines and some nebular lines (e.g. from [NeII] and [SIII]) equivalent widths.

\section{The 12$\mu$m Sample}

Dust absorbs the continuum at short wavelengths and re-emit it in the FIR, however there is a spectral interval (7-12$\mu$m) at which the absorption of the original continuum is balanced by the thermal emission. The 12$\mu$m sample (hereafter 12MGS, Rush, Malkan \& Spinoglio 1993, hereafter RMS)  is an IRAS 12$\mu$m-selected all-sky survey flux-limited to 0.22Jy. From the 12MGS, a relatively unbiased sample of active galaxies in the local Universe has been extracted using optical spectroscopic classification. It is a complete sample in bolometric flux, from the experimental evidence that the flux at 12$\mu$m is about 1/5 of the bolometric one for all types of AGN (Spinoglio \& Malkan 1989; Spinoglio et al. 1995). Moreover, it is less subject to contamination by high star-formation rate objects than other infrared samples defined at longer wavelengths (Hunt \& Malkan 1999) and it is generally used to give the zero point to infrared cosmological studies of galaxies (e.g. Matute et al 2002, Perez-Gonzalez et al 2005).

\begin{figure}[h!]
\resizebox{\hsize}{!}{\includegraphics[clip=true]{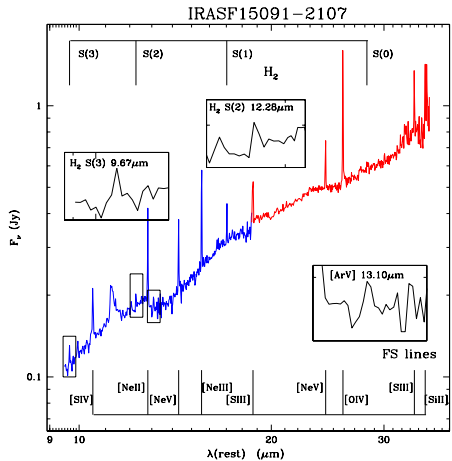}}
\caption{\footnotesize
Example of an high resolution IRS {\it Spitzer} spectrum, of IRAS15091-2107. The PAH features at 11.25$\mu$m and the H$_2$ rotational lines can be seen, as well as the bright mid-IR fine-structure lines originated from either the AGN and from stellar evolution processes. }
\end{figure}

\section{{\it Spitzer} IRS spectra} 

About half (53 objects) of the active galaxies of the 12MGS have been observed by Spitzer IRS at high resolution, in a GT program by G. Fazio and the CfA team in collaboration with us. The data reduction of the first 30 sources observed and their analysis have been reported in Tommasin et al. (2008a, hereafter paper I) and a three-component photoionization model has been computed to separate the emission of 28 sources in three emission line components: Seyfert NLR, starburst and "coronal line region" (Hainline et al. 2007).

The other half of the sample is in the Spitzer public archive and its reduction and analysis are in progress. A second article with other 42 spectra is in preparation (Tommasin et al. 2008b, in prep., hereafter paper II) and a subsequent article will present the overall comparison of data with models (Hainline et al. 2008, in prep.). Fig.1 shows, as an example of the data quality, the IRS spectrum of IRAS15091-2107 (paper I). We anticipate here some of the new results of paper II dealing with 72 objects and therefore with a larger statistics than in paper I. 

\begin{figure}[h!]
\resizebox{\hsize}{!}{\includegraphics[clip=true]{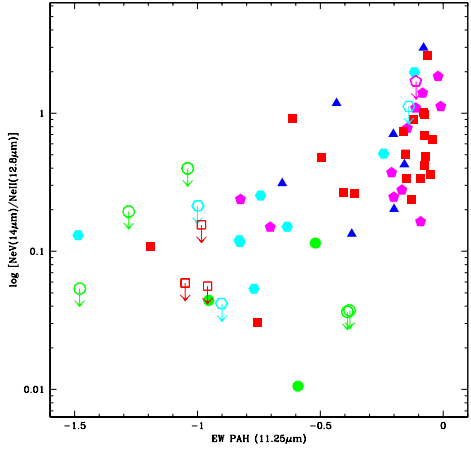}}
\caption{\footnotesize
The PAH 11.25$\mu$m equivalent width as a function of the [NeV]/[NeII] line ratio, which measures the AGN strength. Red squares: Seyfert type 1; Pink pentagons: Seyfert type 2 with hidden BLR; Cyan exagons: Seyfert type 2 without hidden BLR = pure S2; Blue triangles: Seyfert type 2 with no polarimetric data; Green circles: Normal/starburst Galaxy. Full symbols: Detection, Open symbols with arrow: Upper Limit. }
\end{figure}

The ratio of lines of the same element at different ionizing potential gives the galaxy radiation field: [NeV]14.32$\mu$m/[NeII]12.81$\mu$m and [NeIII]15.56$\mu$m/[NeII]12.81$\mu$m are among the prime AGN tracers (Spinoglio \& Malkan 1992, Sturm et al. 2002), while the intensity of the features of the Polyciclyc Aromatic Hydrocarbons are inversely proportional to the AGN activity, as it was suggested from the analysis of  ISO spectra of a sample of ULIRGs  (Genzel et al. 1998). We find the same trend for the Seyfert galaxies of our sample: the PAH equivalent width  decreases as the  [NeV]14.32$\mu$m/[NeII]12.81$\mu$m ratio increases (originating [NeII] at 12.81$\mu$m in the SF regions) (Fig.2). Besides the difference between type 1's and type 2's, in this figure we also distinguish among the Seyfert 2's with a measured hidden broad line region (h-BLR) and those without (pure-Seyfert 2's) and those without polarization measurements. The classification of the Seyfert type 2's in h-BLR and pure-Seyfert 2's, as well as the reclassification of Seyfert galaxies in normal or starburst galaxies, has been taken from Tran (2001, 2003) and Shu et al. (2007) and the former is based on optical spectro-polarization observations. It appears from the figure
that almost all galaxies (23/25) having [NeV]14.3$\mu$m  detections with the \textbar  PAH 11.25$\mu$m \textbar $<$ 0.3 $\mu$m  have a BLR (14 type 1's, 9 h-BLR), while only two objects are "pure-Seyfert 2's" and 4 objects do not have polarimetric data. This is confirmed by a Kolmogorov-Smirnov 2-dimensional test that shows that Seyfert 1's and h-BLR Seyfert 2's are drawn from the same population (with P=0.979).
We suggest that the "pure" Seyfert 2's have on average a lower [NeV]/[NeII] ratio and therefore a weaker AGN component. This result will be further discussed in paper II.

\begin{figure}[h!]
\resizebox{\hsize}{!}{\includegraphics[clip=true]{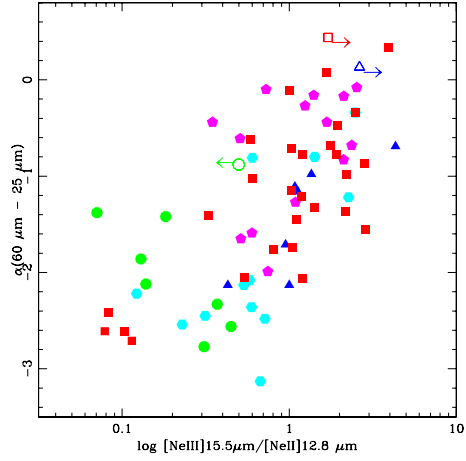}}
\caption{\footnotesize
[NeIII]15.5$\mu$m/[NeII]12.8$\mu$m line ratio versus the (60$\mu$m -25$\mu$m) spectral index. Symbols  as in the previous figure. Most broad-lined objects have a higher [NeIII]/[NeII] ratio and a warmer spectral index than objects without a BLR.}
\end{figure}

As already shown in paper I, we present in Fig.3 the [NeIII]15.5$\mu$m/[NeII]12.8$\mu$m line ratio versus the (60$\mu$m -25$\mu$m) spectral index, with an higher statistics and, as in the previous figure, the separation between Seyfert 2's with hidden broad lines and those without. The ionization in general increases with the flattening of the spectral index.  Using a Kolmogorov-Smirnov 2-dimensional test, we find that the Seyfert type 1 and type 2 should be drawn by the same population (P=0.887). However, Seyfert 1's and h-BLR Seyfert 2's cluster in the upper-right part of the diagram, while most "pure" Seyfert 2's and non-Seyfert's are segregated in the lower-left part, except for 4 objects, classified as type 1's in RMS. One of these is indeed a Seyfert 1.9 (from NED) and the others do not have mid-IR high ionization lines detected, which might question the old optical classification.

\section{X-ray Counterparts}

Objects obscured at soft X-rays but detectable at hard X-rays are defined Compton-thick, as they present a very high absorption hydrogen column density (N$_{H}$ \textgreater 10$^{24}$ cm$^{-2}$) (e.g. Guainazzi 2006). On the other side, objects with a lower N$_{H}$ (\textless 10$^{24}$ cm$^{-2}$) are called Compton-thin. A large fraction of the 12MGS Seyferts are Compton-thick, as can be seen in Fig.4, which plots their IR luminosity versus the hydrogen column density. On the contrary, hard X-rays are not very efficient in detecting Compton-thick objects, especially at high luminosities, as can be seen from the figure which includes the AGN samples observed at hard X-rays by Integral-IBIS and Swift-BAT (Bassani et al. 2006, Markwart et al. 2005). Therefore one of the advantages of the 12MGS is that it allows a combined X-ray/IR study that can help in understanding the role of AGN in the IR galaxy population and the relationship between the accretion power, the obscuration and the star formation properties of the galaxies.

\begin{figure}[h!]
\resizebox{\hsize}{!}{\includegraphics[clip=true]{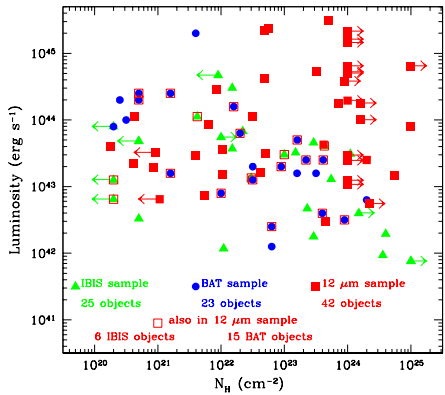}}
\caption{\footnotesize   12$\mu$m Sample Seyferts cover a wide range of hydrogen column density, from Compton-thin objects to Compton-thickÕs, while the X-ray selected sources cover a range.  (N$_{H}$ data from Bassani et al. 2006, Markwart et al. 2005 and Shu et al. 2007)}
\end{figure}

\section{Conclusions}

In Tommasin et al. (2008a) we found no intrinsic IR spectroscopic difference between type 1 and type 2 Seyferts. We confirm this result, however this study, which increases the statistics from 30 to 72 objects, indicates that those Seyfert type 2 that have an hidden BLR, as seen from spectro-polarimetry, lie preferentially in the regions of Seyfert type 1's, for what concerns the AGN dominance and [60 - 25] spectral index,  while "pure" Seyfert 2's have stronger "star formation" components, as measured by the PAH EW.

The fact that our sample of galaxies well covers the plane {\it luminosity-hydrogen column density} containing Compton-thick objects even at high luminosities makes it a suitable sample to study the three parameters governing galaxy evolution: accretion luminosity (and black-hole properties), star formation strength and obscuration. 

\bigskip

\footnotesize {\bf Acknowledgments} This work is based on observations made with the Spitzer Space Telescope which is operated by the Jet Propulsion Laboratory and Caltech under a 
contract with NASA. ST acknowledges support by ASI. We thank Giovanni Fazio and the IRAC Team at CfA for contributing Guaranteed Time to obtaining these data.

\end{document}